\documentclass{aa}
\usepackage{natbib,aas_macros}

\newcommand{\al}{\textsc{AstraLux}}
\newcommand{\minms}{\textsc{MinMS}}

\newcommand{\nbody}{\textsc{Nbody}}

\newcommand{\msun}{M_{\odot}}
\newcommand{\mpc}{M_{\odot} {\rm pc}^{-3}}
\newcommand{\mjup}{M_{\rm Jup}}

\begin{document}

\titlerunning{Predicting the binary fraction in Taurus}
\authorrunning{M. Marks et al.}

\title{Using binary statistics in Taurus-Auriga to distinguish between brown dwarf formation processes}

\author{M. Marks\inst{1}
	  \and
	E. L. Martín\inst{2}
	  \and
	V. J. S. B\'ejar\inst{3,4}
          \and
	N. Lodieu\inst{3,4}
	  \and
        P. Kroupa\inst{5,6}
	  \and
	M. E. Manjavacas\inst{7}
	  \and
	I. Thies\inst{5}
	  \and
	R. Rebolo López\inst{3,4}
	  \and
	S. Velasco\inst{3,4}
}

\institute{Clara-Fey Gymnasium Bonn, Rheinallee 5, 53173 Bonn \email{astro.michi@yahoo.com}
	      \and
	   Centro de Astrobiología (INTA-CSIC), Carretera de Ajalvir km 4, 28550 Torrejón de Ardoz, Madrid, Spain
	      \and
	   Instituto de Astrofísica de Canarias (IAC), Calle vía Láctea s/n, 38200 La Laguna, Tenerife, Spain
	      \and
	   Universidad de La Laguna, Dpto. Astrof\'{i}sica, E-38206 La Laguna, Tenerife, Spain
	      \and
	   Helmholtz-Institut f\"ur Kern- und Strahlenphysik, University of Bonn, Germany
	      \and
	   Charles University in Prague, Faculty of Mathematics and Physics, Astronomical Institute, V  Hole\v{s}ovi\v{c}k\'ach 2, CZ-180~00~Praha~8, Czech~Republic
	      \and
	   Steward Observatory, The University of Arizona, Tucson, AZ 85721, USA
	   }

\date{}

\abstract
{Whether brown dwarfs (BDs) form just as stars directly from the gravitational collapse of a molecular cloud core (``star-like'') or whether BDs and some very low-mass stars (VLMSs) constitute a separate population which form alongside stars comparable to the population of planets, e.g. through circumstellar disk (``peripheral'') fragmentation, is one of the key questions of the star-formation problem.}
{For young stars in Taurus-Auriga the binary fraction has been shown to be large with little dependence on primary mass above $\approx0.2\msun$, while for BDs it is $<10$\%. Here we investigate a case in which BDs in Taurus formed dominantly, but not exclusively, through peripheral fragmentation, which naturally results in low binary fractions. The decline of the binary frequency in the transition region between star-like formation and peripheral formation is modelled.}
{A dynamical population synthesis model is employed in which stellar binary formation is universal with a large binary fraction close to unity. Peripheral objects form separately in circumstellar disks with a distinctive initial mass function (IMF), own orbital parameter distributions for binaries and a low binary fraction according to observations and expectations from smoothed particle hydrodynamics (SPH) and grid-based computations. A small amount of dynamical processing of the stellar component is accounted for as appropriate for the low-density Taurus-Auriga embedded clusters.}
{The binary fraction declines strongly in the transition region between star-like and peripheral formation, exhibiting characteristic features. The location of these features and the steepness of this trend depend on the mass-limits for star-like and peripheral formation. Such a trend might be unique to low density regions like Taurus which host dynamically largely unprocessed binary populations in which the binary fraction is large for stars down to M-dwarfs and low for BDs.}
{The existence of a strong decline in the binary fraction -- primary mass diagram will become verifiable in future surveys on BD and VLMS binarity in the Taurus-Auriga star forming region. It is a test of the (non-)continuity of star formation along the mass-scale, the separateness of the stellar and BD populations and the dominant formation channel for BDs and BD binaries in regions of low stellar density hosting dynamically unprocessed populations.}
\keywords{binaries: general -- stars: late-type -- stars: low-mass -- stars: brown dwarfs -- stars: formation}
\maketitle

\section{Introduction: BD flavours}
\label{sec:intro}
\subsection{Theory: Star-like BD formation}
A pre-stellar core can condense out from a self-gravitating collapsing molecular cloud to form a star \citep{andre2014,lomax2015} when it reaches the density- and temperature-dependent Jeans-mass of the ambient medium \citep{l1998,bb2005ejectedembryo,bcb2006,klessen2007}. Fragmentation of this kind can continue until a core becomes opaque to its own cooling radiation so that the Jeans-mass cannot decrease further \citep{rees1976olf,ll1976olf,silk1977olf}. This is likely the dominant channel to build up the population of hydrogen-burning stars. In his review, \citet{luhman2012} summarizes further mechanisms that have been proposed to also produce brown dwarfs: In the gravitational compression and fragmentation picture, tidal shear and high velocities in a star forming cluster prevent low-mass objects to continue accretion to reach stellar masses \citep{bcb2008fragmentation}. Dynamical interactions among fragments or protostars in a massive core could lead to the ejection of some of them, which prematurely halts their accretion \citep{rc2001ejectedembryo,boss2001ejectedembryo,bbb2002ejectedembryo,bbb2003ejectedembryo,gww2004ejectedembryo,umbreit2005ejectedembryo,bate2011ejectedembryoERR}. Photoionizing radiation from massive OB stars might contribute by removing much of the envelope and disk of a low-mass protostar \citep{hester1996photo,kb2003bdorigin,wz2004photo}. Alternatively, turbulent compression and fragmentation of gas in a molecular cloud produces collapsing cores over a wide range of masses \citep{pn2002,pn2004,hc2008,elmegreen2011}. VLMSs and BDs then arise from the smallest cores. \citet{luhman2012} conclude that VLMSs and BDs can form via the star-like formation mechanism without the need to invoke further formation channels.

\subsection{Theory: BD formation through peripheral fragmentation}
\label{sec:peripheralintro}
In order to form a BD or BD-BD binary directly in the turbulent molecular cloud, a sufficiently high density is required for the proto-BD to be self-gravitating, but at the same time the proto-BD cloud core needs to have a very low mass. Such a combination of conditions seems increasingly unlikely with decreasing mass which is why the theoretically calculated IMF decreases steeply below about $0.1\msun$, in tension with observations \citep[and references therein]{thies2015}. This suggests the need for additional formation channels to produce BDs.

Theory and simulations suggest that it is possible to form planets, BDs and VLMSs through ``peripheral fragmentation'', a term coined by \citet{thies2015} to denote formation channels different from star-like formation, like disk or filament fragmentation.\footnote{In previous work we have also referred to peripheral fragmentation as ``BD-like'' formation. We don't use this term here since BDs likely form both star-like and peripherally and it suggests an answer to the question at hand.} \citet{stamwhit2009discfrag} show from their SPH computations, using sink particles and a smoothing length to mimic realistic behaviour, that massive disks arise frequently, but should be observed infrequently since a high proportion of BDs and VLMSs form rapidly within them through disk fragmentation. These become ejected or are pumped to large eccentricities by the mutual interaction between the multiple fragments that form in a single disk. Through the same mechanism BD-BD binaries form in their disks and might eventually be scattered into the field. They conclude that ``disc fragmentation is a robust mechanism'' and that it ``can produce [...] most of the BD stars, and a significant proportion of the very low-mass hydrogen-burning stars''. \citet{bv2012hybrid} improve and confirm these previous results by self-consistent modelling of compact gaseous clumps, a gas disk and the host star, using a grid-based model. The authors conclude that clump ejection from the disc and, thus, the formation of isolated VLMSs and substellar objects is common. They preliminarily estimate that one in every ten stars ejects an embryo. Such objects do \emph{not} need to contract to stellar densities in order to be ejected into the intra-cluster medium. \citet{bv2012hybrid} explicitely demonstrate that their ejection speeds are low and that they do not differ much from those of YSOs and should remain spatially co-located.

\citet{forgan2013cdps} devise a population synthesis model in which objects form through gravitational instabilities in the periphery of circumstellar disks and subsequent tidal downsizing. The majority of objects forming in their models are BDs. \citet{forgan2015cddyn} build upon the latter model and couple it with an \nbody~integrator in order to follow both the effect of fragment-fragment scattering and their dynamical evolution in clusters up to 10~Myr. They demonstrate, among other results, that a large fraction of such objects become ejected from their disks and populate the cluster field as free-floating objects. \citet{li2015fragmentation} follow a comparable approach by making use of the SPH models of \citet{stamwhit2009discfrag} and a \nbody~technique. They also find that most objects having formed through peripheral fragmentation (henceforth ``peripheral objects'') escape from their host star and additionally investigate the orbital properties of BD-BD binaries formed through capture within the disk. They find a separation distribution peaked between $5-10$~AU (their fig.~11), comparable to observations of the BD population in the Galactic field \citep{burgasser2007rev}.

Recently, \citet{lwh2016isolatedBDformationunlikely} find that according to their SPH computations turbulent flows need to be strongly convergent and having comparable speeds in order to collapse and form a BD in isolation. Thus, they consider it unlikely that a large number of observed isolated BDs have also formed in isolation. They conclude that it is ``difficult to envisage this being the only way, or even a major way, of forming isolated brown dwarfs''. \citet{vorobyov2016ejectedembryo} investigate ejection of gaseous clumps from protostellar disks via many-body interactions in a grid-based model, of which about half are self-bound after ejection. Their computations can lead to wide separation free-floating binary clumps in the brown-dwarf mass range. All ejected clumps are true ejections rather than scattered objects since these have velocities greater than the escape speed from the central star.

\subsection{Observation vs. Theory: Continuous decrease of the binary fraction from stars to BDs}
From the above considerations it follows that
\begin{itemize}
 \item there exist two separate formation channels for free-floating BDs, namely that they can form (i) star-like and (ii) through peripheral fragmentation and subsequent dynamical evolution, and
 \item the formation modes cannot easily be distinguished.
\end{itemize}

Notwithstanding these theoretical findings, the case for a single star-like formation mode for both stars and BDs is often made in the literature. One argument put forward as apparent evidence in favour of a single formation process of stars and BDs is the continuously decreasing trend of binary fraction with decreasing primary mass in the Galactic field over and above the hydrogen-burning mass-limit (HBML) \citep[see Sec.~\ref{sec:sum} in this paper for a discussion of further arguments]{dk2013rev,reipurth2014rev}. However, concluding that this is a general trend being indicative of a single, continuous formation mode must be done with caution since the field population is pieced together from stars of many different ages \citep{mk11}, where massive stars are on average significantly younger than solar-type and VLMSs. Figure~6 in \citet{thies2015} shows how the addition of many cluster populations, which have dynamically processed their binaries internally, lead indeed to the same continuous trend crossing the HBML without steep gradients, although the binary fraction for stars was close to $100$\% initially independently of primary mass (the theoretical population) and included separate populations for stars and BDs.

The trend of the binary fraction in coeval populations, i.e. in star clusters and star formation regions, is likely more informative. But also here the dynamical history needs to be properly accounted for before drawing conclusions. Accordingly, a preferably young region hosting a primordial population is a suitable test-bed for investigating this primary-mass dependency.

In order to address the question which channel dominates the BDs seen in the Galactic field, in young clusters and star-forming regions and how far up in mass peripheral fragmentation, and how far down in mass star-like formation can occur, here the example of the Taurus-Auriga star formation region is used. In particular we shall suggest a means to observationally distinguish between single or multiple formation modes of stars and BDs in this region. Taurus-Auriga is unique given its proximity, youth and low-density and, thus, experienced little dynamical alteration of its binary population. The aim of the present study is to model the declining trend from the high binary fraction of stars to the low binary fractions for BDs in this region using separate formation modes and predict observable features in future surveys of Taurus-Auriga type aggregates.

\section{The stellar and substellar population in Taurus-Auriga}
\label{sec:bintau}

\begin{table*}[t]
\centering
\caption{Taurus multiplicity surveys. Observationally inferred binary fractions are compared to those resulting from the model. For the T Tauri observations the initial model population is used, for the other sub-populations an initial density of $350\mpc$ \citep[as in][]{mk11} for Taurus-Auriga embedded clusters is used to account for little dynamical evolution; see Sec.~\ref{sec:seplate}. We provide predictions for binary-fractions in the separation range $<3$~AU where currently no observations are available.}
\label{tab:surveys}
\begin{tabular}{llllll}
                &                 & \multicolumn{2}{c}{binary-fraction (\%)} & & \\
 mass ($\msun$) & separation (AU) & obs & model & comment & Reference \\
 \hline
 - & $50-5000$ & $64\pm8$ & - & Class 0 protostars & \citet{chen2013tauproto} \\
 \hline
 $0.01-0.1$ & $10^{-4}-3$ & - & $11$ & T Tauri, prediction & -\\
 $0.1-2.0$ & $10^{-4}-3$ & - & $19$ & T Tauri, prediction & -\\
 $0.2-2.3$ & $16-252$ & $37\pm9$ & $28$ & T Tauri & \citet{ghez1993} \\
 $0.5-1$ & $3-1400$ & $55\pm11$ & $59$ & T Tauri; lower bound & \citet{simon1995} \\
 solar & $18-1800$ & $42.5\pm7.9$ & $47$ & T Tauri & \citet{kl1998} \\
 \hline
 $0.01-0.1$ & $10^{-4}-3$ & - & $9$ & model prediction & - \\
 $0.1-2.0$ & $10^{-4}-3$ & - & $18$ & model prediction & - \\
 $0.015-0.14$ & $>$4 & $9^{+10}_{-3}$ & $18$ & VLMOs; $>M5.5$ & \citet{kraus2006tau} \\
 $<0.1$ & $>10$ & $4^{+5}_{-1}$ & $6$ & $>M6$ & \citet{todorov2014tauchamusco} \\
 $0.1-0.3$ & $>10$ & $18^{+8}_{-4}$ & $21$ & $M4-M6$ & \citet{todorov2014tauchamusco} \\
 $0.5-2.0$ & $500-5000$ & $21^{+4}_{-3}$ & $15$ & - & \citet{kraus2009} \\
 $0.25-0.7$ & $3-5000$ & ($64^{+11}_{-9}$) & $50$ & CSF$^1$ & \citet{kraus2011tau} \\
 $0.7-2.5$ & $3-5000$ & ($79^{+12}_{-11}$) & $59$ & CSF$^1$ & \citet{kraus2011tau} \\
 $0.2-3.0$ & $10-1500$ & $26.3^{+6.6}_{-4.9}$ & $38$ & raw & \citet{daemgen2015} \\
 $0.7-1.4$ & all & $62\pm14$ & $84$ & corrected for full range & \citet{daemgen2015} \\
 \hline
 \multicolumn{6}{l}{\scriptsize $^1$CSF denotes that the original paper stated the companion star frequency (number of companions per target) which is larger} \\
 \multicolumn{6}{l}{\scriptsize than the corresponding binary fraction (number of targets with at least one companion) used otherwise throughout this paper.}
\end{tabular}
\end{table*}

The Taurus star-formation region has a distance of $\approx140$~pc \citep{kenyon1994,wichmann1998,torres2012tau} and extends over a region of one hundred square degrees on the sky. The age of its population is $\approx1-2$~Myr, but \citet{daemgen2015} found an additional co-moving sub-population of $\approx20$~Myr. Six distinct clusters of very young stars in Taurus were detected by \citet{gomez1993} to have projected radii of $0.5-1$~pc with $\approx15$ binaries each, with a mass of about $6\msun$ if the average stellar mass is $\approx0.4\msun$, and a median separation between the stellar centre-of-mass systems within each cluster of $0.3$~pc. The individual groups are separated by a few pc, thus do not interact dynamically. All of Taurus has more than $300$ known members and the stellar density in each of its groups is, with $1-10$~stars~pc$^{-3}$, very low \citep{luhman2009}. The N-body models of \citet{kb2003tau} show that Taurus-Auriga like arrangements are largely unevolved and in particular that the binary properties should be similar to those at birth, with some dynamical evolution \citep{mk11}.

Taurus exhibits a paucity of high-mass young stars \citep{kenyon2008taurev} and was originally proposed to be deficient in low-mass stars \citep{briceno2002} as well. Many brown dwarfs have however since been discovered \citep{luhman2004T,guieu2006}. \citet{kroupa2003bd} and \citet{tk2008tau} show that there is no need to invoke a non-canonical IMF in Taurus.

\citet{todorov2014tauchamusco} searched for companions to young brown dwarfs in Taurus and Chamaeleon~I. Resolving binaries with separations larger than $10$~AU, they find, combining their results with the studies of \citet{kraus2006tau}, \citet{konopacky2007tau} and \citet{kh2012sfr}, a binary fraction of $18^{+8}_{-4}$\% for M4-M6 type binaries ($0.1-0.3\msun$) and $4^{+5}_{-1}$\% for binaries in their sample with spectral type $>$M6 ($<0.1\msun$) in Taurus. Given its youth and low-density, \citet{todorov2014tauchamusco} suggest that the observed reduction of binary fraction towards later spectral-types might be primordial.

Among $45$~observed T~Tauri stars in Taurus, \citet{ghez1993} find $22$~binaries with angular separations between $16-252$~AU. The binary fraction is thus $\approx49$~percent over this separation range. \citet{simon1995} find $22$ binaries and $4$ triples among $47$~systems. The binary frequency in the range $3-1400$~AU is at least $1.6\pm0.3$~times the value of the canonical Galactic field sample \citep{DuqMay91,r2010}. \citet{kraus2009} find $27$ wide binaries in Taurus in the separation range $500-5000$~AU with masses down to $\approx0.1\msun$, which is consistent with a log-flat distribution but inconsistent with the Galactic field log-normal distribution. \citet{lz1993tau} and \citet{kl1998} identified in total $74$~binaries or multiples among $174$~systems. The binary fraction of $42.5\pm4.9$~percent is larger by a factor of $1.93\pm0.26$ than that of solar-type main sequence stars over the accessible separation range. In a high resolution imaging study \citet{kraus2011tau} find that $2/3-3/4$ of all Taurus stars are multiple. In the most recent near-infrared imaging search for stellar and substellar companions \citet{daemgen2015} target 64 Taurus members with masses between $0.2$~and $3\msun$, identifying secondaries down to $2\mjup$. Within their 90\% completeness limit they find a raw multiplicity fraction of $26.3^{+6.6}_{-4.9}\%$ in the separation range $10-1500$~AU. After completeness correction for the full separation range they find $62\pm14$\% of all Taurus companions in the range $0.7-1.4\msun$ to be multiple, comparable to \citet{kraus2011tau} results. This corresponds to a $\approx1\sigma$ detection for a $1.4\times$ larger binary fraction of solar-type stars in Taurus compared to the Galactic field \citep{r2010}. The frequency of proto-stellar binaries appears to be even larger still \citep{chen2013tauproto}.

All surveys combined thus suggest the stellar binary fraction to be significantly larger in the Taurus-Auriga groups than in the field. We summarize the here mentioned surveys in Table~\ref{tab:surveys}.

\section{Modelling the (sub-)stellar single and binary population of Taurus}
In contrast to Taurus' stellar population, its BDs exhibit a low binary fraction \citep{todorov2014tauchamusco} and we assume that peripherally formed BDs to have a field-like narrow orbital-parameter distribution. A low binary fraction as well as the narrow separation distribution for BDs also results from SPH models of peripheral fragmentation \citep[see Sec.~\ref{sec:peripheralintro}, e.g.][]{li2015fragmentation}. In turn, there will be a strong decline in binary fraction from M-dwarfs to BDs, \emph{independently} of the following model details. The aim here is not to present a final word, but instead to get an idea of how the binary fraction might shrink in this mass range and to predict features in its primary mass dependency if BDs and stars have their own IMFs and binary distribution functions.

\subsection{The IMF}
\label{sec:imf}
\begin{figure}
 \includegraphics[width=0.48\textwidth]{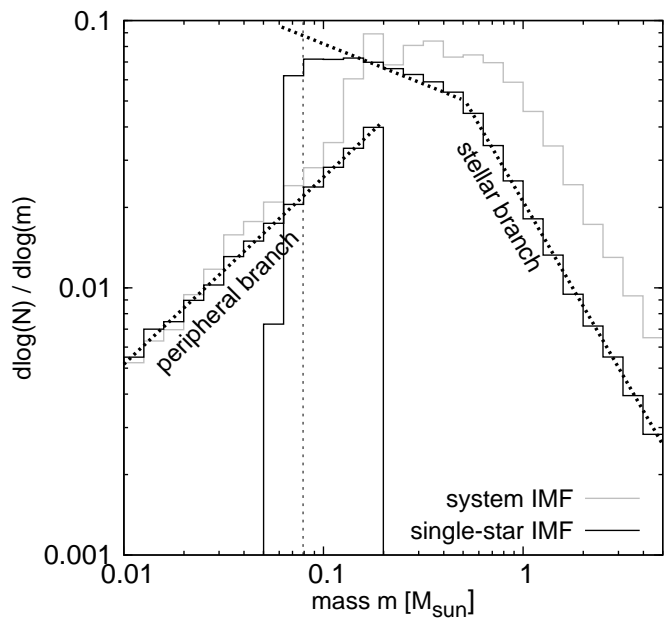}
 \caption{The stellar and substellar IMF are disjoint in the present model (black histograms). The dotted lines indicate an underlying power-law of the mass spectrum the form $dN/dm\propto m^{-\alpha}$ with $\alpha=0.3,\;1.3$ and $2.3$ (from left to right). Objects from the peripheral branch of the IMF lying above the HBML (vertical dashed line) are stars forming in circumstellar disks and star-like BDs stem from the stellar branch of the IMF lying below the HBML. Both populations mix and form the system IMF ($31$\% and $\approx100$\% binary fraction for the BD and stellar branch, respectively, see Sec.~\ref{sec:imf}). The discontinuity is thus hidden from an observer (grey histogram). Figure taken from \citet{marks2015}. \label{fig:imf}}
\end{figure}
In order to model Taurus' (sub-)stellar population, in which BDs and BD binaries form dominantly through peripheral fragmentation and capture in the circumstellar disk, we invoke a combined IMF for BDs and stars which is discontinuous around the HBML but whose distinct branches overlap somewhat in mass \citep[Fig.~\ref{fig:imf}, see also][]{k2013rev}.
$$\xi(m)=\frac{dN}{dm}\propto m^{-\alpha_i}{ },{\rm where} \begin{cases}\alpha_0=0.3 & m<0.2\msun \\ \alpha_1=1.3 & 0.06<m<0.5\msun \\ \alpha_2=2.3 & m>0.5\msun \end{cases}$$
Upon combining both branches, their distinctness disappears in observations, strongly declining near the HBML instead for the here chosen model parameters. The parameters of the IMF in Fig.~\ref{fig:imf} have been obtained empirically by comparing the discontinuous model to observations but can vary somewhat from region to region \citep{tk2007,tk2008tau}.

A sharp cut in mass between the two branches seems odd, since there is no reason why the star-like and peripheral formation mode should care about a common limit, so there is likely some overlap in mass for star-like BD formation and peripheral fragmentation. In the SPH computations of \citet{stamwhit2009discfrag} VLMSs form through peripheral fragmentation up to $\approx0.2\msun$. By comparing the present IMF model to observations of the IMF in Taurus, IC348, the Pleiades and the Orion Nebula Cluster, \citet{tk2007,tk2008tau} constrain the maximum mass for peripheral fragmentation to be close to this mass as well. And \citet{marks2015} require the same limit to model the late M-dwarf binary population in the Galactic field when using separate formation modes. Thus, these independent theoretical and empirical constraints suggest the limit for peripheral fragmentation to lie near $0.2\msun$. The lower limit for star-like formation is here chosen to be $0.06\msun$, a value close to but below the HBML to allow for some BDs to be formed star-like. Using this choice of mass-limits reproduces constraints of Galactic field binary populations \citep{marks2015}.

The discontinuity in the IMF depends on the fraction of peripheral objects to star-like objects in the population, ${\cal R}_{\rm pop}=N_{\rm peripheral}/N_{\rm star-like}$, and is chosen to be $0.3$. This reflects the empirically determined value for Taurus-Auriga and the Pleiades \citep{tk2007}, i.e. for every three star-like bodies there is one object forming through peripheral fragmentation. For the here used parameters 64\% of all BDs in the range $0.01-0.08\msun$ form in circumstellar disks while the rest forms star-like \citep{thies2015}. \emph{Note that these numbers are sensitive to the chosen limits for star-like formation and peripheral fragmentation.}

The IMF parameters should ideally be treated as free parameters, as \citet{tk2007,tk2008tau} have done for individual regions. However, the aim here is not to constrain these parameters, but to model the decline of the binary fraction with decreasing primary mass, for which data are not yet fully available. While a wide range in mass of Taurus members \emph{is} probed by existing studies (see Sec.~\ref{sec:bintau}), the individual analyses cover a limited separation range, being restricted by the used technique, and most studies do not extrapolate their binary fractions for the full separation range. Thus, it is not meaningful to plot the available data in a single diagram and infer a trend from it. Since the BD binary fraction is significantly lower than the stellar binary fraction, varying the IMF parameters does not strongly influence the resulting features in the binary fraction -- primary mass distribution, just their location, as we will qualitatively address in the discussion (Sec.~\ref{sec:sum}).

\subsection{The separation distribution}
\label{sec:sep}
\begin{figure}
 \includegraphics[width=0.48\textwidth]{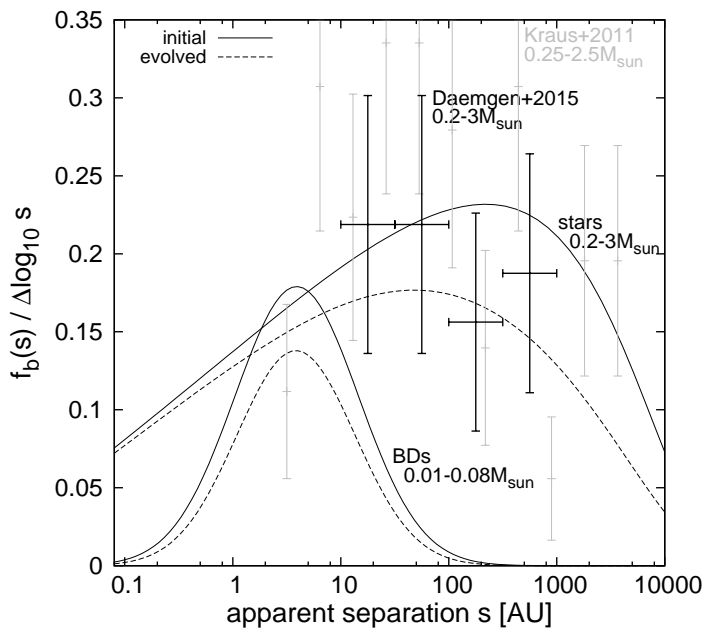}
 \caption{Separation distribution for stars and BDs in the model (curves, Sec.~\ref{sec:sep}) in comparison with observational data on stellar binarity in Taurus \citep[symbols with $\sqrt N$ errors,][]{kraus2011tau,daemgen2015}. \citet{daemgen2015}'s data is constructed using the physically bound pairs among 64 targets associated with Taurus (their table~2, see Sec.~\ref{sec:bintau}). The solid curves show the initial BD and stellar late-type population in the model. The dashed curves depict the resulting distributions after allowing for little evolution in the Taurus-Auriga groups (see Sec.~\ref{sec:sep}). The BD distributions result from the addition of BDs from both IMF branches (Fig.~\ref{fig:imf}). \label{fig:separation}}
\end{figure}

\subsubsection{Late-type stars and star-like BDs}
\label{sec:seplate}
\citet{daemgen2015}'s data is consistent with both a log-flat distribution suggested in previous multiplicity studies \citep{kraus2011tau}, and the shape (not the multiplicity) of the solar-type separation distribution \citep{r2010} in the accessible range of separations ($\approx10-1000$~AU). The latter consistency might be due to the inclusion of a 10$\times$ older sub-population identified by \citet{daemgen2015} which might have undergone more dynamical modification than the $1-2$~Myr old population which likely resembles more closely the one at birth. But also \citet{kraus2011tau} found that lower mass Taurus members ($0.25-0.7\msun$) show a paucity of binary companions with separation $\gtrsim200$~AU, while solar-type stars and above do not. But this may as well be due to the lack of sensitivity to lower mass companions.

The universal\footnote{For a discussion of the concept of Universality see \citet{marks2015}.} birth binary population chosen here for all late-type stars and star-like BDs in Taurus is the one of \citet[his equation~8]{k95b},
$$\Phi^{\rm birth}_{\log_{10} P}(m\lesssim2\msun)=2.5\frac{\log_{10}P-1}{45+\left(\log_{10}P-1\right)}\quad,$$
which has a maximum period $\log_{10}P_{max}=8.43$, where P is measured in days, and a $100$\% initial binary fraction. The very long maximum birth orbital period is a formal solution and corresponds to a binary separation of about $0.03$~pc for a system mass of $1\msun$. This is the dimension of about a molecular cloud core and therefore constitutes a physically plausible case. Such systems will be easily torn apart through dynamical interactions in dense environments, but are required to match observations in young extended regions and star forming regions. Comparably wide systems are observed in Taurus (Fig.~\ref{fig:separation}).

This birth population is subjected to pre-main sequence eigenevolution, a process capturing the effects in the gas-embedded phase of proto-binary clumps, to yield the initial population. The initial population is the one before any dynamical modification takes place. Since the observed period distributions of G-, K- and M-dwarfs are indistinguishable in shape, we assume that BDs selected from the star-like branch of the IMF, which formed through primary fragmentation like stars, have the same initial distribution function as G-, K- and M-dwarfs.

While it may be argued that there are many different possible choices for the initial binary period distribution function, the one derived by \citet{k95b} was based on an iterative procedure in which solutions to both the Galactic-field late-type distributions and simultaneously to the pre-main sequence Taurus-Auriga constraints were sought, subject to stellar-dynamical modifications of these distribution function which were taken care of using state-of-the-art \textsc{Nbody} computations. Consistency with the observed distribution of specific angular momenta in pre-stellar cloud cores was sought as well. Later work also demonstrated consistency with the binary populations in the dynamically evolved Orion Nebula Cluster and the Pleiades cluster \citep{kah2001oncpleiades,mk12,marks2014}, in the Galactic field \citep{mk11,marks2015} as well as with trends of binary fraction in present-day globular clusters \citep{leigh2015bingc}.

As we can see from Fig.~\ref{fig:separation}, both \citet{kraus2011tau}'s and \citet{daemgen2015}'s data is compatible with this initial population. To account for the older sub-population we allow for some dynamical modification of the initial population over $5$~Myr by placing it inside a typical Taurus-like sub-cluster, applying an analytical description of dynamical processing of binary populations in young star clusters.

Since the binding-energy of a binary is supposedly key to whether a binary can be dissolved by dynamical interactions, \citet{mko11} devise a time-dependent stellar-dynamical operator, $\Omega^{M_{ecl},r_h}_{\rm dyn}(t)$, which acts on the initial binding-energy distribution, $\Phi_{\log_{10} E_b,in}$, of primordial binaries, which results from \citet{k95b}'s initial separation distribution and random pairing of masses from the IMF. This operator modifies the original binding-energy distribution to a dynamically evolved one,
$$\Phi^{M_{ecl},r_h}_{\log_{10} E_b}(t)=\Omega^{M_{ecl},r_h}_{\rm dyn}(t)\times\Phi_{\log_{10} E_b,in}\quad,$$
from which other processed distributions, such as the separation and mass-ratio distribution, can be extracted via a Monte-Carlo method. In this formulation,
$$\Phi^{M_{ecl},r_h}_{\log_{10} E_b}(t)=\frac{f_b(\log_{10} E_b,t)}{\Delta\log_{10} E_b}\quad,$$
is the time-dependent binary fraction, $f_b=N_b/N_{\rm cms}$, i.e. the number of binaries per targets (``center-of-mass'' objects) normalized to the log-width of the binned distribution. A similar definition holds for the separation distribution in Figure~\ref{fig:fb}. The parameters describing the stellar-dynamical operator are found by following the evolution of binary populations in \nbody-computations of clusters of different initial embedded stellar mass, $M_{ecl}$, radius, $r$, and, thus, density. \citet{mko11} find that the initially binary-dominated population changes on a crossing-time scale, i.e the resulting distributions after a few Myr depends only on the initial stellar density, the population should not be strongly altered beyond that time.

Upon applying the operator for a moderate initial density \citep[$350~\mpc$, see][]{mk12} on the initial distributions for Taurus' age of $\approx1$~Myr, we see that the agreement with \citet{daemgen2015}'s data persists for the processed initial population.

\subsubsection{Peripheral Objects}
\label{sec:sepbd}
Each BD binary and VLM M-dwarf binary selected from the peripheral branch is assigned a semi-major axis selected from the observed narrow separation distribution of BDs in the Galactic field. The used Gaussian distribution has a width $\sigma=0.4$ (in $\log_{10}$~AU units) with a peak at $\log_{10}s/{\rm AU}=0.66$ \citep[according to][using the data from the Very Low Mass Binary Archive]{tk2007}. Most of the selected objects remain single, though, to meet Galactic field constraints upon combining them with star-like BDs. \citet{marks2015} empirically determine the total binary fraction for binaries on the IMF peripheral branch to be $\approx31$\%, assuming the Galactic field BD population has been largely unaffected by dynamical processing in their progenitor embedded clusters.

We expect this to be largely the case since if BD binaries form dominantly through peripheral fragmentation there will be a natural truncation of the separation distribution at large separations due to the extent of circumstellar disks of $\lesssim$few~hundred~AU. This leaves BD binaries with separations which are not strongly affected by dynamical processing in host clusters they were born in, at least up to moderate initial densities. Note that the fraction of BD binaries in the separation distribution can decrease nevertheless due to dynamical disruptions of BD binaries and stellar binaries with a BD companion originating from the star-like branch, as this adds single BDs (dashed BD distribution in Fig.~\ref{fig:separation}).

\section{Results: The primary-mass dependent binary fraction in between the mass-limits for star-like and peripheral formation}
\label{sec:binfrac}
\begin{figure*}
\centering
\includegraphics[width=0.48\textwidth]{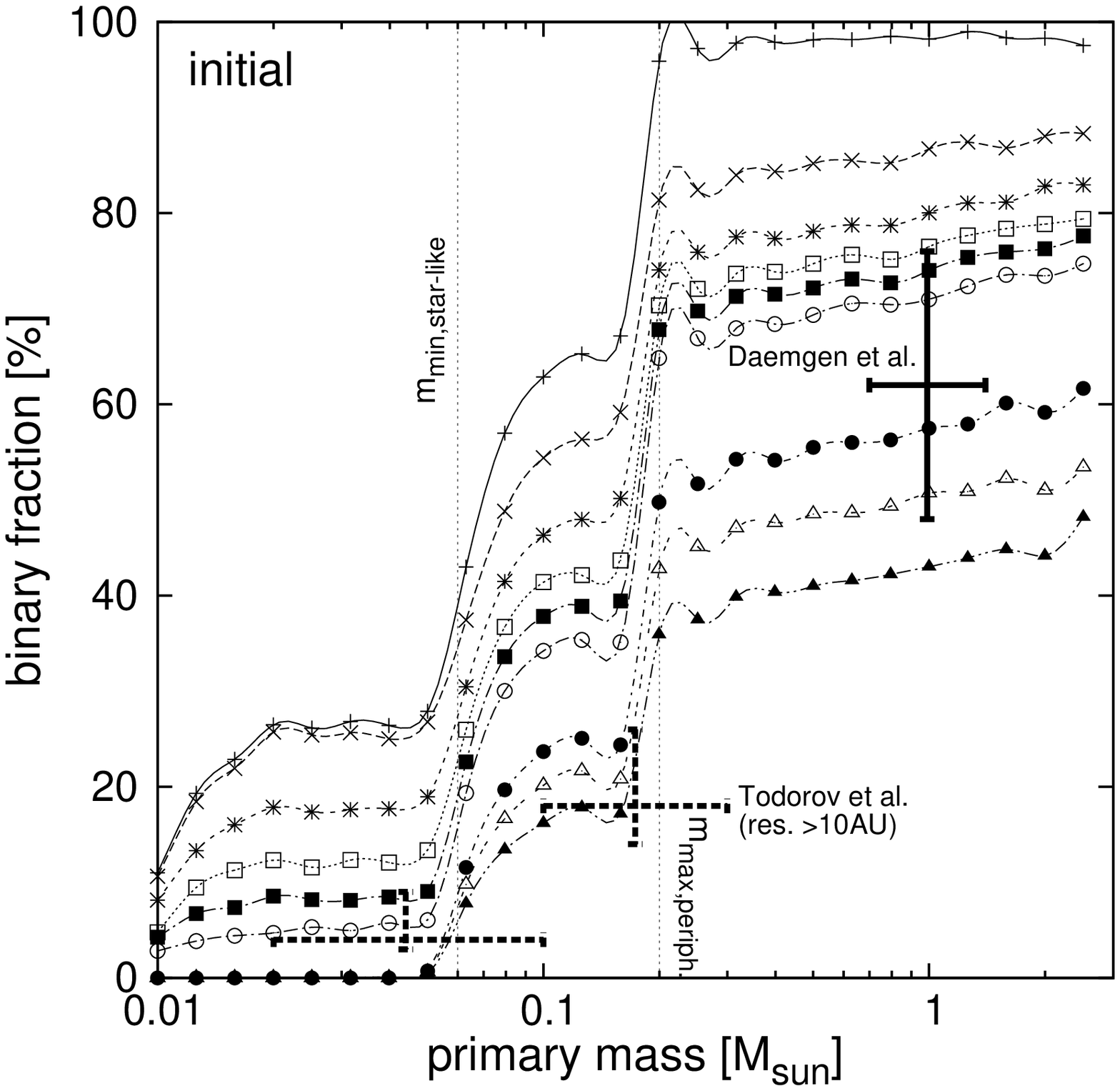}
\includegraphics[width=0.48\textwidth]{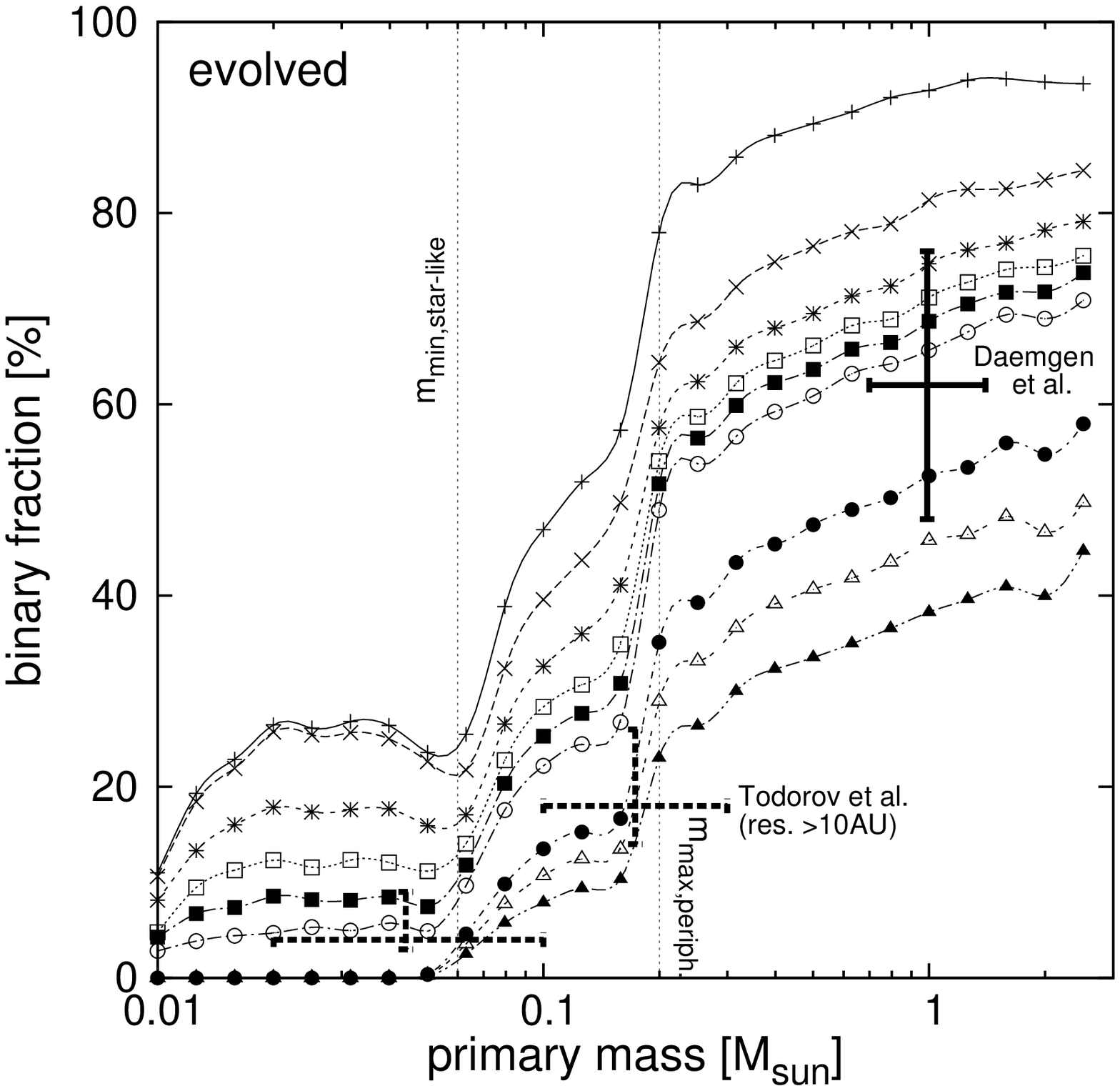}
\caption{Prediction for the primary-mass dependent binary fraction in dependence of resolution limit, if peripheral fragmentation applies (the discontinuous IMF model in Fig.~\ref{fig:imf}). \emph{Left:} Binary fraction as a function of primary mass for the initial binary population with no dynamical evolution (Sec.~\ref{sec:sep}). The different curves denote the trend for varying survey resolution limits: Binaries with separations larger than 0, 1, 3, 5, 7, 10, 50, 100 and 200~AU (top to bottom) are resolved. Binaries with separations smaller than the respective resolution limits are counted as single in the model. The solid cross denotes the \citet{daemgen2015} estimation that $62\pm14\%$ of solar-type Taurus members over the full separation range are multiples, in concordance with \citet{kraus2011tau} finding that about 2/3 $-$ 3/4 of their targets in Taurus are binaries or higher order multiples. The dashed crosses are data from \citet{todorov2014tauchamusco} who resolve separations $a>10$~AU. \emph{Right:} Same as left panel but with dynamical evolution in the Taurus-Auriga groups added (see Sec.~\ref{sec:sep}). \label{fig:fb}}
\end{figure*}

\citet{daemgen2015} find that the binary fraction of \emph{young} Taurus members shows no correlation with mass in the range $0.2-3\msun$, contrasting the \emph{older} subpopulation for which they find $\approx3\sigma$ evidence for a declining trend with decreasing primary-mass. If it doesn't have a primordial origin, this age dependency might suggest stellar dynamical processes to be at work. Combining both populations results in a weak $\approx1\sigma$ trend. \citet{daemgen2015} show their results for Taurus to compare well to earlier results obtained for Upper~Sco \citep{lafreniere2014usco}.

In Fig.~\ref{fig:fb} (left panel) the trend of binary fraction as a function of primary mass in the model predicts the same absence of a trend for young stars in Taurus if dynamical evolution has no strong impact. Furthermore the binary fraction for stars should decline weakly at best towards lower masses until peripheral objects enter the binary statistics (here below $0.2\msun$, cf. Fig.~\ref{fig:imf}) the binary fraction strongly declines due to the sudden presence of many single objects having formed peripherally in circumstellar disks. Another dip occurs when the star-like formation mode ceases to exist (here at $0.06\msun$, cf. Fig.~\ref{fig:imf}) and peripheral fragmentation is the only remaining formation channel available to BDs and planets below that mass. The right panel of Fig.~\ref{fig:fb} shows the same, but with dynamical modification of the initial population in a small Taurus-like cluster with an initial density of $350\msun/pc^3$ \citep{mk12}. The overall binary fraction has dropped a little due to the break-up of binaries. A weak correlation of binary fraction with increasing primary mass shows up as well for hydrogen-burning stars since lower mass binaries have on average a lower binding energy and are more susceptible to dissolution. This trend is as well qualitatively consistent to \citet{daemgen2015}'s results. The features occuring at the depicted mass limits for star-like formation and peripheral fragmentation are otherwise similar. The observed binary fraction depends on the resolution limit of the survey.

The observational constraints on the binary fraction by \citet{todorov2014tauchamusco} and \citet{daemgen2015} agree somewhat better with the models that allow for dynamical processing of the binary population in Taurus. In particular the \citet{todorov2014tauchamusco} survey data which resolved binaries with semi-major axes $>10$~AU is consistent with the curve representing the model with a resolution limit $>10$~AU.

Inclusion of other observational studies into Fig.~\ref{fig:fb}, as discussed in Sec.~\ref{sec:bintau}, is not reasonable since a relatively limited separation range is covered in comparison. Instead we compare the observationally inferred binary fractions to the model by subjecting the model to the separation and mass constraints of the Taurus surveys listed in Tab.~\ref{tab:surveys}. The model yields very good agreement to within the observational errors (column 3 vs. 4) for the T Tauri surveys as well as for VLMSs. For higher-mass stars in Taurus the model binary fractions lie inside 2$\times$ the observational error bars. In terms of binary fractions the model is thus overall consistent with the available surveys.

For the separation range $0-3$~AU currently no observation is available. To be tested against future observations, our model predicts (see Tab.~\ref{tab:surveys}) a binary fraction lying near $\approx10$~per~cent for BD binaries in the mass-range $0.01-0.1\msun$ within this separation range, and $18-19$~per~cent for binaries with a stellar primary in the mass-range $0.1-2.0\msun$, both for TTauri~stars as well as for Taurus' older subpopulation.

\section{Discussion}
\label{sec:sum}

\subsection{Distinguishing BD flavours in Taurus-Auriga}
The present analysis has demonstrated how the large binary fraction for stars in Taurus-Auriga is expected to decline steeply from M-dwarfs towards BDs if stars form during the collapse of a proto-stellar cloud and BDs form preferentially, but not exclusively, through peripheral fragmentation in circumstellar disks. From our modelling we do not expect a rather smooth decrease as might be inferred from the available observational data alone. The strong decline in a relatively small mass-range determined by the mass limits for peripheral and star-like formation is equally consistent with available data.

The characteristic features in Fig.~\ref{fig:fb} depend on the mass-limits for star-like and peripheral formation. Mass-limits lying further apart might flatten the trend in between these limits, while close by mass-limits would pronounce the decline. A decline near these mass-limits with a ``saddle'' in between, where stellar and peripheral branch overlap, is characteristic in either case. While the upper mass-limit for peripheral formation of $0.2\msun$ is constrained by different studies (see Sec.~\ref{sec:imf}) and expected to lie close to this value in Taurus, the lower mass-limit for star-like formation is rather arbitrarily chosen here. Incoming observational data will help to better constrain the values for $m_{\rm min,star-like}$ and $m_{\rm max,peripheral}$ in Taurus-Auriga, if these mass-limits are considered as free parameters.

We note that, while the characteristic trend shown in Fig.~\ref{fig:fb} is specific to the present model, the general decline seen here for Taurus-Auriga like groupings should be independent of the model details, if only BDs form dominantly through peripheral fragmentation, i.e. have a formation channel of their own which produces BD binaries much less frequent than the star-like formation mode and the stellar population has been unaffected by dynamical processing of its binary population. Further observables discussed here such as the separation distribution are model specific and can test the idea that the BD separation distribution is different from the one for stars.

If the here predicted behaviour for the primary-mass dependent binary fraction near the HBML were confirmed in future observations, this would then be evidence for peripheral fragmentation being dominant for BD production in Taurus-Auriga, and possibly in other regions of low stellar surface density as well. That is, if other hitherto unknown scenarios in which BD binaries form star-like \emph{and} have a low binary fraction can be excluded. We're not aware of a mechanism that would cause a similar decline in the binary fraction. Qualitatively, the existence of a critical angular momentum below which binary formation is hindered could be an option. \citet{machida2008ISMfragmentation} demonstrate using MHD nested grid simulations that fragmentation in a collapsing cloud is controlled by the initial ratio of the rotational and magnetic energy, where rotation promotes and magnetic fields hinder fragmentation. If this were the mechanism to produce BD binaries much less frequent than stellar binaries, it remains qualitatively unclear why fragmentation into binaries suddenly becomes much less efficient close to the HMBL.

\subsection{And what about dense star formation regions?}
In denser regions disks might disrupt quickly due to tidal interactions. The frequency of disks around stars is found to decrease with age \citep{haisch2001discs,muzerolle2010discs,hernandez2008discs,mamajek2009discs}, being consistent with a dynamical reduction, and questions whether peripheral fragmentation can be dominant in dense environments. Thus, generalizing a positive result for Taurus-Auriga to regions of higher stellar density would, in this picture, be possible only if the frequency of disk fragmentation (which may depend on the mass infall rate onto the disk, thermal physics in the disk, etc.) is larger than the frequency of disk-destructive encounters. On the other hand, the increased rate of stellar encounters may also lead to more triggered BD formation \citep{thies2005}.

\citet{pfalzner2014discs} caution that disk lifetimes might be considerably longer than inferred from a disk frequency -- age trend since the above observational studies are biased towards the innermost parts of (formerly denser) clusters while the outskirts of clusters or those systems which nowadays populate the Galactic field might have been exposed to very different, less destructive conditions.\footnote{After all, the majority of field stars investigated by the \textsc{Kepler} space-telescope are known to harbour planets. How would this be possible if circumstellar disks of now planet-harbouring stars are destroyed in their parent clusters?}

For the above reasons, peripheral fragmentation might still have been the dominant channel in regions of higher stellar density and, thus, for the BD population that nowadays exists in the Galactic field.

\subsection{Ambiguity of observational tracers of continuous star formation}
\label{sec:ambigious}
According to theory (Sec.~\ref{sec:intro}), BDs can likely form both star-like and through peripheral fragmentation. However, the community favours continuous star formation along the mass-scale, i.e. a single formation mode, in explaining their observations. In the following we discuss tracers that are frequently said to favour a common, single formation mode for stars and BDs.

\subsubsection{Continuity of the IMF at the VLMS -- BD transition}
\citet{km2012imf} see a decline in the number of objects per $\log(M/\msun)$ at about $0.1$ in the differential mass distribution of the nearby star-forming regions Taurus, Lupus3, ChaI and IC348, which persists for stars both in regions of high and low stellar density (their figure~12). In others, like the $\sigma$~Orionis open cluster \citep{bihain2009sorionis,caballero2007sorionis,lodieu2009sorionis,penaramirez2012sorionis} and in the Pleiades \citep{bihain2006pleiades,zapatero2014pleiades} a decline is not obvious near the HBML. All mass distributions appear to show a continuous transition from BDs to stars, albeit exhibiting different gradients. Does this evidence continuity in the star formation process?

\citet{tk2007} show that a proper treatment of BD multiplicity properties and unresolved binarity among stars reveals a discontinuity in observed continuous mass distributions, which, in contrast, indicates separate formation modes. The discontinuity remains pronounced when larger binary fractions among BDs are allowed for \citep{tk2008tau}. A recent analysis by \citet{thies2015} furthermore shows that analytical theoretical derivations of the IMF, which assume BDs to be a continuous extension to stars, underestimate the numbers of observed BDs and thus require a correction term to account for the BD population, again interpreted to favour different formation channels.

Thus, whether or not a mass distribution seems to be continuous observationally, and whether it declines (strongly) near the HBML or not, it can be modelled with a distinct BD and stellar branch of the IMF and is thus \emph{not} an unambigious tracer of star-like formation.

\subsubsection{Spatial distributions and velocities of VLMSs and BDs in very young open clusters}
A similar spatial distribution for stars and BDs is observed in $\sigma$~Orionis \citep{caballero2008sorionisspacialdist,lodieu2009sorionis}, in $\rho$~Ophiuchus \citep{parker2012rhooph} and in the 25~Orionis group \citep{downes2014sorionisspacialdist}. Such observations are interpreted as being indicative of a single formation mode.

However, SPH computations of peripherally formed BDs show that depending on whether already completely formed BDs or their gaseous progenitors are ejected or scattered, they can reach different escape speeds from their discs. As already discussed in Sec.~\ref{sec:peripheralintro}, the hybrid grid-based model of \citet{bv2012hybrid} demonstrated that ejection speeds of peripherally formed objects in their computations are low and that they do not differ significantly from the relative velocities of young stellar objects. In turn, BDs and stars should remain spatially co-located within their scenario, that is if later mass segregation through cluster dynamical evolution is either unimportant or the cluster is sufficiently young. That implies that co-location is no conclusive indicator of a single formation mode for both stars and BDs.

\subsubsection{Molecular outflows, accretion disks and disk frequencies around (isolated) BDs}
It is often argued that observations of outflows, BD disks and their high disk frequencies demonstrate that sub-stellar objects form as a simple continuation of the stellar process.

The faint object SMM2E will most likely remain a substellar object \citep[$30-35~M_{\rm Jupiter}$,][]{palau2014bd}, just as a $73~M_{\rm Jupiter}$ object in Taurus \citep{phanbao2014bd}. These authors interpret their respective objects to be a scaled-down version of a forming star. Even the free-floating planetary mass object OTS44 ($12~M_{\rm Jupiter}$) is interpreted to form star-like based on the existence of a substantial disk and significant accretion \citep{joergens2014fflop}.

Disks around VLMSs and BDs are observed to be more frequent than around higher-mass stars in $\sigma$~Orionis \citep{luhmann2008sorionisdiskfreq,penaramirez2012sorionis}, $\lambda$~Orionis \citep{bayo2012lorionisdiskfreq}, the Upper Scorpius association \citep{luhman2012uscodiskfreq,lodieu2013usco} and the nearby young cluster NGC~1333 \citep{scholz2012ngc1333}. The opposite trend is seen in Chamaeleon~I \citep{luhman2008chamdiskfreq,luhman2010taudiskfreqERR}.

As disks around later type stars are exposed to a less destructive radiation field, the authors ascribe their findings to different disk life-times for different stellar masses, indicating a continuous transition from stars to BDs. This finding is equally consistent with higher-mass stars losing their disks much quicker than lower-mass stars through gravitational interactions in higher-density clusters \citep{pfalzner2006diskdispersal}. In this picture, the disk fractions of stars will have been higher in the past, perhaps even overhauling the disk fractions of BDs.

However, if BDs (or isolated planets) form in the disks of proto-stars through either encounter-triggered perturbations \citep{thies2010bd} or gravitational instabilities \citep{stamwhit2009discfrag,bv2012hybrid,forgan2013cdps} and are nudged out or ejected, the resultant free-floating objects will carry with them their own mass reservoir. Using their SPH computations, this has explicitly been shown by \citet{stamwhit2009discfrag} and \citet{thies2015}. As a result, all available models of BD formation are consistent with the detection of BD disks, a conclusion which is shared by \citet{luhman2012}. In turn, observations of isolated BDs and massive planets carrying their own mass reservoir, having outflows and exhibiting a large disk frequency among them is no evidence in favour of or against a particular formation mode.

Recently, \citet{ricci2014bddisks} observe extended disks to three BDs ($>70$~AU) which are not easily reconciled in a pure peripheral fragmentation scenario according to which BD disk sizes are limited to $<20$~AU \citep{bate2009a,bate2012sph}. If true, this does however not exclude the here presented IMF model with separate branches for peripheral and star-like objects, respectively, since star-like BDs, which may have more extended disks, are not excluded in our model. On the other hand the work by \citet{bate2009a,bate2012sph} is confronted by the study of \citet{vorobyov2016ejectedembryo} suggesting extended disks to exists for (at least some) ejected proto-BDs, because the ejected clumps often possess high angular momentum.

\subsubsection{Width of the separation distributions of Galactic field binaries}
\label{sec:bingf}
The binary-fraction and width of the observed log-normal separation distributions in the Galactic field decline continuously with decreasing mass of the primary component \citep{DuqMay91,fm1992m,d2004m,bergfors2010m,r2010,janson2012m,jodar2013km,tok2014fg,dk2013rev,reipurth2014rev,wardduong2015,cortes2017m}. This continuity across the mass-limit between stars and BDs was apparently supported by recent investigations of the late M-dwarf binary population in the Galactic field. In a lucky-imaging survey, the \al~survey for earlier \citep{janson2012m} and later \citep{janson2014m} M-dwarfs (the two studies overlap somewhat in mass) showed separation distributions significantly narrower than that for solar-type stars, confirming results from a previous lucky-imaging survey \citep{bergfors2010m}. For the late M-dwarfs the separation distribution lies remarkably close to the one for BDs in the Galactic field, suggesting a common origin. The conclusion that the distribution for M-dwarfs in the lucky imaging results is indeed that narrow originates from the observation that the distribution starts to decline strongly far from the resolution limit of the respective surveys.

This apparent continuity might however be the result of the density-dependent dynamical alteration of an environment-independent birth binary population for late-type stars in young star forming regions \citep{k95b,k2011,kp2011,mk12,marks2014} and the subsequent addition of many such populations originating from different environments that comprise the Galactic field \citep{pgkk2009,g2010,mk11,marks2015}. In particular, the mutual proximity of BD and late-M separation distributions in \citet{janson2014m} might be following from separate BD and stellar populations which underwent dynamical processing, as \citet{marks2015} have demonstrated. As for the early M-dwarf data, \citet{marks2015} pointed out that the lucky-imaging results are potentially in tension with the M-dwarfs in Multiples \citep[\minms,][]{wardduong2015} data, which covers a comparable spectral-type range, but finds a separation distribution for early M-dwarfs that has approximately twice the width inferred from the surveys of \citet{bergfors2010m} and \citet{janson2012m}. A wider distribution would be in-line with the distributions found by other previous studies \citep{fm1992m,d2004m}.

The notion of continuous star formation over and above the HBML from present Galactic field data could thus be a result of dynamical processing in the host clusters from which the field population originated, hiding any underlying separate (primordial) binary populations for stars and BDs. Observations noting the shrinking width of separation distributions in the Galactic field (but in star formation regions as well) can't be explained unequivocally via a continuous star formation process.

\subsection{Distinguishing BD flavours elsewhere}
\label{sec:distinguish}
Next to our suggestion here to probe BD formation in Taurus-Auriga, three other recent studies try to shed light on how to decide whether star-like or peripheral formation is dominant in producing BDs.

\begin{itemize}
 \item[1.] \citet{stam2015discs} suggest that for peripheral objects there is no (significant) correlation between the mass of the object and their disks or the accretion rates from the disk onto it. This would deviate from a suggested linear disc mass -- stellar mass dependence and corresponding accretion rates onto star-like formed objects derived for masses $\gtrsim0.2\msun$ \citep[with considerable scatter though,][]{mohanty2013discscale,andrews2014discscale} and is thus a potential diagnostic to distinguish different BD formation modes.
 \item[2.] \citet{marks2015} argue that a separate peripheral population might unmask itself through a peak on-top of a wider separation distribution for late M-dwarfs ($\lesssim0.2\msun$) in dynamically \emph{processed} populations if some VLMSs form through peripheral fragmentation with Galactic field BD properties.
 \item[3.] While the hitherto mentioned observables focus on the properties of populations of BDs to distinguish formation channels, they cannot decide whether an individual (isolated) BD formed peripherally or star-like. \citet{vorobyov2016ejectedembryo} suggests a potential means to decide the formation mode on an object-by-object basis. From his grid-based computations the internal structure of ejected, thus free-floating, gaseous pre-BD clumps might be different from those formed through molecular cloud fragmentation. The former are more centrally condensed and have a higher central temperature than pressure-supported, gravitationally contracting spheres. Also, the rotational velocity of ejected clumps has a bi-modal distribution since they are often rotationally supported, making them very distinct from the rotational pattern of star-like forming objects.
\end{itemize}

\section{Summary}
The question investigated here is how to distinguish whether BDs form dominantly via peripheral fragmentation in circumstellar disks or star-like in Taurus-Auriga. We have argued that both formation channels are possible according to theory (see Section~\ref{sec:intro}), which is why our models allow for both star-like and periphal formation of BDs. Peripheral formation is allowed down to $0.06\msun $and star-like formed BDs are allowed to range up to $0.2\msun$ for various reasons (see Section~\ref{sec:imf}), i.e. both channels overlap somewhat in mass.

We find that the binary fraction as a function of primary mass in Taurus-Auriga is expected to decline steeply in a narrow mass-range between the mass limits for the star-like and peripheral mode according to our models with separate formation channels. It exhibits characteristic features in dependence of these limits (see Figure~\ref{fig:fb} and Section~\ref{sec:binfrac}). If observed in future surveys this could be interpreted as evidence for peripheral fragmentation being dominant for BD production in Taurus-Auriga. Such a trend might be unique to low stellar surface brightness (low density) regions hosting dynamically largely \emph{unprocessed} populations.

Our models predicts a binary fraction of $\approx10$~per~cent and $18-19$~per~cent for primary masses in the range $0.01-0.1\msun$ and $0.1-2.0\msun$, respectively, for binaries with separations $<3$~AU. Since no observation for this separation range is currently available, these numbers can be tested against future observations as well.

We note that peripheral fragmentation produces critical constraints matched by observations, e.g. the shape of the IMF around the HBML \citep{tk2007,tk2008tau,thies2015}, the binary statistics of BDs which are different from higher mass stars \citep[Sec.~\ref{sec:bingf}]{tk2008tau,dieterich2012}, the generation of free-floating planetary mass objects and BDs \citep{stamwhit2009discfrag,li2015fragmentation,forgan2015cddyn,vorobyov2016ejectedembryo} and the lack of BD companions to solar-type stars at low separations known as the BD desert \citep{marcybutler2000desert,grether2006desert,dieterich2012,evans2012,wilson2016desert}. All these observables are expected from peripheral fragmentation models and are thus no unambigious tracers of a single star-formation mode (see Sec.~\ref{sec:ambigious}). Instead, we put forward further potential actual tracers of BD formation in Section~\ref{sec:distinguish}.

Concluding, probing the binary population observationally in the transition region between star-like and peripheral formation ($\approx0.05-0.3\msun$) in Taurus and other low stellar surface brightness star-forming regions might be a valuable effort to come closer to answering questions about the (non-)continuity of star formation across the VLMS/BD mass-range, the separateness of the stellar and BD populations and the dominant formation channel for BDs and BD binaries. It is important, though, that our information is not too restricted by separation ranges, so that surveys covering different mass ranges become directly comparable at best, i.e. by combining various methods. If these are not available we encourage authors to try and estimate binary fractions and orbital parameter distributions for the whole separation range, if reasonable.

\section*{Acknowledgements}
This paper originated from a discussion between PK and the IAC brown dwarf group during PK's visit in June 2015, and we thank Alejandro Vazdekis for funding this visit. EM is supported by grant AYA2015-69350-C3-1-P and NL and VJSB are supported by grant AYA2015-69350-C3--2-P both from the Spanish Ministry of Economy and Competitiveness (MINECO).

\bibliographystyle{aa}
\bibliography{biblio.bib}
\makeatletter   \renewcommand{\@biblabel}[1]{[#1]}   \makeatother

\clearpage

\end{document}